\documentclass[journal=jacsat,manuscript=article]{achemso}
\setkeys{acs}{articletitle=true}
\setkeys{acs}{keywords = true}

\usepackage[version=3]{mhchem} 
\usepackage{comment}
\usepackage{cases}
\usepackage{amsmath}
\usepackage{graphicx}
\usepackage{lineno}
\usepackage{color}
\usepackage{soul}
\usepackage{subcaption}

\author{Federico Cao}
\affiliation{Department of Applied Mathematics, University of Waterloo, Waterloo, Ontario, Canada}
\author{Dmitry Eskin}
\affiliation{Skolkovo Institute of Science \& Technology, Moscow, Russia}
\author{Yuri Leonenko}
\affiliation{Department of Earth and Environmental Sciences, University of Waterloo, Waterloo, Ontario, Canada}
\altaffiliation{Department of Geography and Environmental Management, University of Waterloo, Waterloo, Ontario, Canada}
\email{leonenko@uwaterloo.ca}

\title{Modeling of Ex-Situ Dissolution for Geologic Sequestration of Carbon Dioxide in Aquifers}

\abbreviations{CO_2, CCS, GHG, IPCC}
\keywords{ex-situ dissolution; geologic sequestration; modeling}
\begin{document}

\begin{abstract}
Underground carbon dioxide ($CO_2$) sequestration is considered to be one of the main methods to mitigate greenhouse gas (GHG) emissions. In this technology, pure $CO_2$ is injected into an underground geological formation and since it is less dense than residual fluids, there is always a risk of leakage to the surface. To increase security of underground $CO_2$ disposal, ex-situ dissolution can be implemented. When $CO_2$ is dissolved in brine before injection, it significantly reduces the risks of leakage. In this approach, pure $CO_2$ is dissolved on the surface before injection. Surface dissolution could be achieved in a pipeline operating under the pressure of a target aquifer into which the $CO_2$ is injected. In a pipeline, $CO_2$ droplets are dissolved being dispersed in a brine turbulent flow. 
In this paper, a comprehensive model of droplet dissolution along a pipeline is presented. The model accounts for droplet breakup and coalescence processes and is validated against available experimental data.
\end{abstract}

\section{Introduction}

As rapidly developing economies require higher energy consumption, it is clear that major greenhouse gas (GHG) emitting sources cannot be avoided in future decades. Currently, there are serious limitations to alternative/sustainable energy sources as they are still cost-prohibitive for many industries and in developing countries. As a result, fossil fuel consumption will continue to be the main source in the near future. Annual global carbon emissions from fossil fuels have increased to nearly 10 billion metric tons in 2014 (\citeauthor{USEnergy}, \citeyear{USEnergy}).
With $CO_2$ as the most common GHG and responsible for 65\% of anthropogenic global warming, it is crucial to determine feasible mitigation measures. In particular, the Intergovernmental Panel of Climate Change (IPCC) reports that carbon capture and storage (CCS) methods can be an effective solution in significantly lowering the amount of $CO_2$ in the atmosphere (\citeauthor{IPCC}, \citeyear{IPCC}).

Various CCS technologies exist; however, $CO_2$ sequestration in sedimentary basins is of particular interest among many. This form of sequestration relies upon depleted oil and gas reservoirs, (\citeauthor{Herzog2001}, \citeyear{Herzog2001} and \citeauthor{Jenkins2012}, \citeyear{Jenkins2012}) unmineable coal bed reservoirs (\citeauthor{Shi2005}, \citeyear{Shi2005}) and deep aquifers (\citeauthor{Celia2015}, \citeyear{Celia2015}) where the saline water (brine) is not suitable for agricultural or  consumption purposes. Another sequestration option is ocean storage, where $CO_2$ would be injected into the ocean at depths of over one thousand meters (\citeauthor{Haugan2004}, \citeyear{Haugan2004}). Among the above options, deep saline aquifers represent the largest long term potential for CCS (\citeauthor{IEA}, \citeyear{IEA}). An IPCC special report 
(\citeauthor{IPCC}, \citeyear{IPCC}) has suggested that deep saline formations have a storage capacity of around 2000 Gigatons (Gt) of $CO_2$. It is approximately two orders of magnitude higher than the total annual worldwide emissions amount, making saline aquifers the most viable disposal option. 
Although it was recognized that deep saline aquifers offer very large potential storage capacity, significant uncertainties remain regarding storage security. $CO_2$ injected into a saline aquifer is less dense than the resident brine and, driven by buoyancy, will flow horizontally, spreading under the cap-rock which should confine $CO_2$ for thousands of years until it is fully dissolved.  Cap-rocks of aquitards have not been proven to hold buoyant $CO_2$ for geologic time scales as in the case for cap-rocks that have confined buoyant oil and gas (\citeauthor{Van1993}, \citeyear{Van1993}; \citeauthor{Linde1997}, \citeyear{Linde1997}). It also may flow upward, leaking through any high permeability zones such as natural fractures or artificial penetrations such as abandoned wells. Therefore, approaches which allow an increase of storage security are of great importance for developing and implementing CCS technologies.
One of the ways to increase storage security is to accelerate dissolution of $CO_2$ in a formation brine and several methods have been reported in the literature. After $CO_2$ is injected into a reservoir it starts to dissolve naturally (natural convection). Both the onset of convection and the dissolution rate are strongly dependent on reservoir parameters, and an overview of these phenomena is studied in detail in the work of \citeauthor{Emami2015} (\citeyear{Emami2015}). The time scale for complete dissolution, in this case, could  reach  $10^3-10^5$ years. To speed up the dissolution process a variety of engineering options has been proposed. For example, \citeauthor{Hassan2009} (\citeyear{Hassan2009}) and \citeauthor{Keith2005} (\citeyear{Keith2005}) suggested a method for enhancing $CO_2$ dissolution in saline aquifers by injecting brine on top of the injected $CO_2$. This technique significantly increases the dissolution rate, and half of the injected $CO_2$ is dissolved within 200 years. To further improve the dissolution process, employment of mixing devices, such as a static mixer (\citeauthor{Zirrahi2013a}, \citeyear{Zirrahi2013a}) or a downhole mixer (\citeauthor{Zirrahi2013b}, \citeyear{Zirrahi2013b}) has been studied. \citeauthor{Shafaei2012} (\citeyear{Shafaei2012}) investigated using an injection well of a special design, where brine and $CO_2$ are co-injected. Those devices have the potential to significantly increase the dissolution rate. In the current study, we investigate a possibility of nearly $100\%$ dissolution before it is injected underground. 

\begin{figure}
    \centering
    \includegraphics[width=0.6\textwidth]{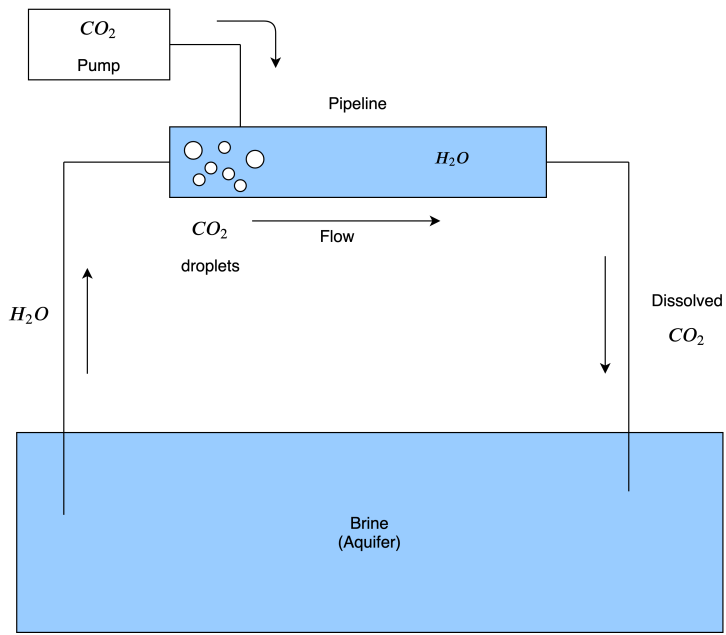}
    \caption{A representation of ex-situ dissolution. Captured $CO_2$ is mixed with brine along a turbulent pipeline flow. The $CO_2$ saturated brine is then injected underground into saline aquifers.}
    \label{fig:ESD}
\end{figure}

The idea was first proposed in our previous study (\citeauthor{Leo2008}, \citeyear{Leo2008}), where we adopted the view that the only relevant risk of leakage arises from mobile free-phase $CO_2$, which is not immobilized by residual or chemical trapping or dissolution. Therefore storage security mainly depends on two factors: (a) the likelihood that free-phase $CO_2$ will leak out of the reservoir and (b) the rate at which free-phase $CO_2$ is immobilized by one of the trapping mechanisms. Storage security then can be increased either by reducing the probability of leakage or by increasing the rate at which $CO_2$ is immobilized by residual gas trapping, dissolution in reservoir fluids, or geochemical reactions. In the same study we proposed some options to reduce the time scale of free phase of $CO_2$: in-situ and ex-situ dissolution. The latter could be achieved within a surface pipeline where two phase $CO_2$-brine mixture flow takes place. The generation of $CO_2$ droplets, which are sufficiently small to achieve rapid dissolution, occurs in a turbulent pipe flow. We present a diagram of the ex-situ dissolution procedure in Fig.~\ref{fig:ESD}.

In our former studies, (\citeauthor{Zendehboudi2011}, \citeyear{Zendehboudi2011}; \citeauthor{Cholewinski}, \citeyear{Cholewinski}) mass transfer from $CO_2$ droplets into brine during co-current pipeline flow was modeled to investigate the effectiveness of the proposed method. The models, however, did not include droplet breakup and coalescence. In a subsequent study (\citeauthor{Zendehboudi2013}, \citeyear{Zendehboudi2013}), a very simplistic model of droplet breakup was employed where coalescence was entirely ignored.
Let us briefly discuss the flaws of the model of \citeauthor{Zendehboudi2013} (\citeyear{Zendehboudi2013}). Other mentioned prior models are based on an even larger number of assumptions. First, the model of \citeauthor{Zendehboudi2013} (\citeyear{Zendehboudi2013}) describes mean droplet diameter evolution along a pipe, whereas a real dispersed system of droplets is strongly polydispersed. Second, this model assumes a constant turbulence energy dissipation rate over a pipe cross-section, whereas it is maximum at the pipe wall and very strongly decreases toward the pipe center. The energy dissipation rate is the predominant driver of all the dispersion processes considered. Third, the assumption made by \citeauthor{Zendehboudi2013} (\citeyear{Zendehboudi2013}) that mass transfer between a droplet and an ambient fluid can be calculated through turbulent diffusion flux is incorrect. \citeauthor{Levich} (\citeyear{Levich}) showed that the diffusion flux is calculated through the mean relative droplet-fluid velocity, caused by turbulent fluctuations. 
The dispersion-dissolution model proposed in the present work is free of the aforementioned drawbacks because it takes into account droplet polydispersity, non-uniformity of flow parameters across a pipe cross-section and employs a validated empirical correlation to calculate the mass transfer coefficient for a droplet moving in a turbulent flow.

Thus, the model of \citeauthor{Zendehboudi2013} (\citeyear{Zendehboudi2013}) could be considered only as an engineering estimation of the dissolution process, whereas the model we developed represents a mathematical description of the complex dissolution process based on first principles.

In the current paper we perform comprehensive modeling of ex-situ dissolution by incorporating all three phenomena which take place in a pipeline: droplet coalescence, breakup, and dissolution.
Dispersion of droplets is modeled by an advection-diffusion population balance equation. The numerical results obtained are validated versus the available experimental data.

\section{Modeling}
To simulate dispersion of $CO_2$ droplets in the pipeline, let us formulate the major assumptions:
\begin{enumerate}
\item Droplet dispersion over a pipe cross-section is intense enough to neglect gravity induced droplet stratification.
\item Flow is steady-state.
\end{enumerate}
Assumption 1 appears to be reasonable taking into account that the density of liquid carbon dioxide for typical aquifer conditions ($700-800$ kg/m$^3$) is not significantly different from that of water. Furthermore, $CO_2$ droplets are rapidly fragmented in a turbulent pipe flow down to small sizes. Due to turbulent diffusion, such droplets are rather uniformly distributed across a pipe even if it is horizontal. Also, dissolution significantly contributes to a droplet size reduction, additionally supporting the validity of the latter assumption. We would like to emphasize that we calculated a steady-state hydrotransport of particles in a water pipe flow by the model of \citeauthor{Eskin2012} (\citeyear{Eskin2012}) assuming that the particle density is equal to that of liquid $CO_2$. For droplets of size $150$ $\mu$m transported at a holdup $0.1$ in a pipe of the diameter $0.15$ m with the mean velocity $3$ m/s, a droplet volume concentration variation across a pipe turned out to be only about $20$\%. This variation is relatively small, whereas even initially large droplets in a turbulent pipe flow are rapidly dispersed (\citeauthor{Eskin2017}, \citeyear{Eskin2017}) reaching the aforementioned size of $150$ $\mu$m; therefore, assumption 1 is acceptable.\\
Assumption 2 is straightforward because we consider steady-state operational conditions with a constant flow rate.\\
Let us formulate the advection-diffusion-population balance equation for a steady-state pipe flow in cylindrical coordinates for a dispersed phase discretized by $M$ size fractions as follows (\citeauthor{Eskin2017}, \citeyear{Eskin2017}):
\begin{linenomath}
\begin{align} \label{eq:1}
u(r)\frac{\partial{N_i}}{\partial{x}} &= \frac{1}{r} \frac{\partial}{\partial{r}}\left(rD_i(r)\frac{\partial{N_i}}{\partial{r}}\right) + u(r)\left(\frac{\partial{N_i}}{\partial{x}}\right)_{PB}, \hspace{.5cm} i = 1,...,M.
\end{align}
\end{linenomath}
Here, $i$ is the size fraction number, $r$ is the radial coordinate, $u(r)$ is the flow velocity, $N_i$ is the number concentration of droplets of the $i$-th size fraction in a computational cell, $D_i(r)$ is the turbulent diffusivity of a droplet of the $i$-th size fraction, and $\left(\dfrac{\partial{N_i}}{\partial{x}}\right)_{PB}$ is the number concentration derivative accounting for coalescence, breakup and dissolution for the $i$-th size fraction.
The boundary conditions for Eq.~\eqref{eq:1} are formulated as
\begin{enumerate}
\item The volume flux through the pipe wall is zero:
\begin{equation}
q_i(R) = -D_i \dfrac{\partial{N_i}}{\partial{r}}\Bigr{|}_{r=R} = 0.
\end{equation}
\item The dispersed phase concentration gradient at the pipe axis is zero: 
\begin{equation}
\dfrac{\partial{N_i}}{\partial{r}}\Bigr{|}_{r=0} = 0.
\end{equation}
\item The droplet size distribution at the initial pipe cross-section is: \begin{equation}
N_i(0,r) = \Psi_i(r)
\end{equation}
where $\Psi_i(r)$ is a some known function.
\end{enumerate}

\noindent
Eq.~\eqref{eq:1} can be rewritten in a one-dimensional time-dependent form as follows:
\begin{equation}
\dfrac{u(r)}{U}\dfrac{\partial{N_i}}{\partial{t}} = \dfrac{1}{r}\dfrac{\partial}{\partial{r}} r \left(D_i(r) \dfrac{\partial{N_i}}{\partial{r}}\right) +  \dfrac{u(r)}{U}\left(\dfrac{\partial N_i}{\partial t}\right)_{PB} \label{eq:1DTD}
\end{equation}
where $U$ is the mean pipe flow velocity and $dt = dx/U$.

\noindent
Further, we need to describe physical processes accompanying droplet size evolution in a pipe.
Fortunately, dispersed pipeline flows have been rather intensely studied
in the past; therefore, for our modelling we will use an engineering approach based on mainly validated ideas. The following sections describe models employed to calculate the terms of Eq.~\eqref{eq:1DTD}.

\subsection{Flow velocity distribution across a pipe}
The steady-state velocity distribution across a pipe $u(r)$ is assumed to be consisting of the two regions:
1) the viscous layer in the wall vicinity, characterized by a linear velocity distribution; 2) the turbulent boundary layer, extended from the viscous layer to the pipe center and characterized by a logarithmic velocity distribution. In
dimensionless coordinates, this velocity distribution is written as (\citeauthor{Gersten}, \citeyear{Gersten}):
\begin{linenomath}
\begin{align}
    u^{+} &= y^+, \hspace{.5 cm} y^+ \leq 11.6 \label{veldist1} \\
    u^+ &= 2.5 \ln y^+ + 5.5, \hspace{.5 cm} y^+ > 11.6 \label{veldist2} 
\end{align}
\end{linenomath}
where $u^+ = u/u_{*}$ is the dimensionless flow velocity, $u_{*} = (\tau_w/\rho_f)^{0.5} $ is the friction velocity, $y^+ = u_{*}y/\nu_f$ is the dimensionless coordinate, $y = R-r$ is the distance from the wall, $\nu_f$ is the fluid kinematic viscosity, and $\tau_w$ is the wall shear stress.

\subsection{Turbulent diffusivity of droplets}
The droplet turbulent diffusivity can be determined as: 
\begin{linenomath}
\begin{align}
    D_p = \dfrac{D_t}{Sc_{pt}}
\end{align}
\end{linenomath}
where $D_t = v_t/Sc_t$ is the turbulent diffusivity, $v_t$ is the eddy diffusivity, $Sc_t$ is the Schmidt number for a fluid particle in a turbulent flow, and $Sc_{pt}$ is the turbulent Schmidt number for a particle (droplet).\\
In the present work, to determine the droplet turbulent diffusivity distribution across a pipe, we will assume that the droplet turbulent diffusivity is equal to the eddy diffusivity:
\begin{linenomath}
\begin{equation}
    D_i = v_t. \label{eq:turbed} 
\end{equation}
\end{linenomath}
Since for the dispersion system considered in the present work droplet sizes as well as a difference between densities of a fluid and a dispersed phase are relatively small, Eq.~\eqref{eq:turbed} is justified.\\
The dimensionless eddy diffusivity distribution across a pipe can be calculated using the empirical
equations suggested by \citeauthor{Johansen} (\citeyear{Johansen}):
\begin{linenomath}
\begin{align}
    v_t^+ &\approx \dfrac{v_t}{\nu_f} = \left(\dfrac{y^+}{11.15}\right)^3  \hspace{0.5 cm} \text{for} \hspace{0.5 cm} y^+ \leq 3\\
    v_t^+ &\approx \left(\dfrac{y^+}{11.4} \right)^2 - 0.049774 \hspace{0.5 cm} \text{for} \hspace{0.5 cm} 3 < y^+ \leq 52.108\\
    v_t^+ &\approx \kappa y^+ \hspace{0.5 cm} \text{for} \hspace{0.5 cm} y^+ > 52.108
\end{align}
\end{linenomath}
where $\kappa = 0.406$ is the von Karman constant.

\noindent
The derivative, expressing the population balance term in Eq.~\eqref{eq:1DTD} can be represented as a sum of the derivatives determining the contributions of breakup, coalescence and dissolution respectively:
\begin{linenomath}
\begin{align} \label{eq:totalpop}
    \left(\dfrac{\partial N_i}{\partial t} \right)_{PB} = \left(\dfrac{\partial N_i}{\partial t} \right)_{break} + \left(\dfrac{\partial N_i}{\partial t} \right)_{coal} + \left(\dfrac{\partial N_i}{\partial t} \right)_{diss}.
\end{align}
\end{linenomath}
For numerical calculation of both breakup and coalescence terms, we employed the Fixed Pivot Approach of
\citeauthor{Kumar1996} (\citeyear{Kumar1996}).

\subsection{Breakup term}
For droplet breakup modeling, we employ a usual binary breakup assumption: fragmentation of a mother droplet leads to formation of two daughter droplets.\\
The derivative associated with droplet breakup is calculated as
(\citeauthor{Kumar1996}, \citeyear{Kumar1996}): 
\begin{linenomath}
\begin{align} \label{eq:br}
    \left(\dfrac{\partial N_i}{\partial t}\right)_{break} = \sum_{k=1}^M n_{i,k}G_kN_k - G_iN_i
\end{align}
\end{linenomath}
where $G_i$ is the breakup rate of a droplet belonging to the $i$-th size fraction (a model will be given further). The function $n_{i,k}$ is calculated as follows: 
\begin{linenomath}
\begin{align}
    n_{i,k} = \int_{x_i}^{x_{i+1}} \dfrac{x_{i+1} - v}{x_{i+1}-x_i}\beta(v,x_k) dv + \int_{x_{i-1}}^{x_i}\dfrac{v-x_{i-1}}{x_i - x_{i-1}} \beta(v,x_k)dv
\end{align}
\end{linenomath}
where $\beta(v,x_k)$ is the droplet breakup density function characterizing the probability of formation of a droplet of the $k$-th size fraction of the volume $x_k$ at the breakup of a droplet of volume $v$. Both first and second integrals for $i=1$ and $i=k$ are zero  respectively.\\

\noindent
To use \eqref{eq:br}, the equations of both the droplet breakup rate and breakup density function
should be specified. In the present research, we will use the functions employed by \citeauthor{Eskin2017} (\citeyear{Eskin2017} and \citeyear{Eskin2017TC}) in their work on modeling droplet dispersion in a pipe. The equation for breakup rate at a low droplet concentration is as follows:
\begin{linenomath}
\begin{align}\label{eq:G_d}
    G(d) = K \dfrac{(\epsilon d)^{1/3}}{d}\left[erfc(\Phi^{1/2}) + \dfrac{2}{\pi^{1/2}}\Phi^{1/2}\exp(-\Phi) \right]
\end{align}
\end{linenomath}
where $erfc$ is the complimentary error function, $K$ is the model parameter, $\epsilon$ is the energy
dissipation rate per unit mass, $\Phi = \dfrac{3}{2}\dfrac{We_{cr}}{We}$ is the dimensionless parameter,
$We = \dfrac{2\rho_f (\epsilon d)^{2/3} d}{\gamma}$ is the Weber number for a droplet, $We_{cr}$ is the critical Weber number (experimental parameter).\\
\citeauthor{Eskin2017TC} (\citeyear{Eskin2017TC}) identified $We_{cr}$ from the Couette device experiments as $We_{cr} = 0.5$. This parameter determines the steady-state droplet size distribution. Due to a very rapid dispersion process in a Couette device, the parameter $K$, determining the rate of a size distribution change, was not identified accurately and was assumed to be $K = 1$ (\citeauthor{Eskin2017TC}, \citeyear{Eskin2017TC}). In the present work, we will employ the critical Weber number recommended by \citeauthor{Eskin2017TC} (\citeyear{Eskin2017TC})  and allow the parameter $K$ to be varied to fit the $CO_2$ droplet dissolution experimental data. We will also employ the same breakup density function as \citeauthor{Eskin2017} (\citeyear{Eskin2017}) used in their research on droplet dispersion in a pipe:
\begin{linenomath}
\begin{align}
    \beta(f_{bv}) = 12 f_{bv}(1-f_{bv})
\end{align}
\end{linenomath}
where $f_{bv} = v/x$ is the breakup fraction, $x$ and $\beta$ are the mother and smaller droplet volumes respectively.
According to \citeauthor{Eskin2017TC} (\citeyear{Eskin2017TC}), this function weakly affects particle size distributions calculated by solving the population balance equation (Eq.~\eqref{eq:br}).\\

\noindent
The energy dissipation rate for the employed velocity profile (Eq.~\eqref{veldist1}, \eqref{veldist2}) is calculated as the specific power spent on friction between concentric fluid layers (\citeauthor{Eskin2017}, \citeyear{Eskin2017}):
\begin{linenomath}
\begin{align}
    \epsilon(\tilde{r}) = \xi \dfrac{\tilde{r}^{3/2}}{1-\tilde{r}}
\end{align}
\end{linenomath}
where $\tilde{r} = r/R$ is the dimensionless radius, $\xi = (-0.5 \nabla p/\rho_f)^{1.5}\sqrt{R}/\kappa$ is the dimensional complex, $\nabla p$ is the pressure gradient, $\kappa$ is the von Karman constant.\\
The pressure gradient in a pipe flow is calculated as:
\begin{linenomath}
\begin{align}
    \nabla p = -2\rho_f f U^2/D.
\end{align}
\end{linenomath}
In the present work, we will calculate the Fanning friction factor $f$ by the Blausius equation, valid for hydraulically smooth pipes, as (e.g., \citeauthor{Bird}, \citeyear{Bird}):
\begin{linenomath}
\begin{align}
    f = \dfrac{0.079}{\text{Re}^{0.25}}
\end{align}
\end{linenomath}
where Re $ =UD/\nu_f$ is the pipe Reynolds number.  

\subsection{Coalescence term}
The derivative we used to account for coalescence in Eq.~\eqref{eq:totalpop} is calculated as (\citeauthor{Kumar1996}, \citeyear{Kumar1996}):
\begin{linenomath}
\begin{align}
    \left(\dfrac{\partial N_i}{\partial t}\right)_{coal} = \sum_{\substack{j,k \\ x_{i-1} \leq x_{j}  + x_{k} \leq x_{i+1}}}^{j \geq k} (1 - 0.5 \delta_{jk}) \eta Q_{j,k}N_jN_k  - N_i \sum\limits_{k = 1}^M Q_{i,k} N_k 
\end{align}
\end{linenomath}
where $\delta_{jk}$ is the Kronecker delta function and $Q_{j,k}$ is the coalescence rate of droplets belonging to the $j$ and $k$ size fractions.\\
The variable $\eta$ is calculated by the following equations: 
\begin{linenomath}
\begin{numcases}{\eta  = }
\dfrac{x_{i+1} - v}{x_{i+1} - x_i}, & $x_i \leq v \leq x_{i+1}$ \\
\dfrac{v - x_{i-1}}{x_i - x_{i-1}}, & $x_{i-1} \leq v \leq x_i$
\end{numcases}
\end{linenomath}

\noindent
There are many models for calculation of the coalescence rate, which are available in the literature (\citeauthor{Yixing2010}, \citeyear{Yixing2010}). However, literature analysis (\citeauthor{Yixing2010}, \citeyear{Yixing2010}) shows that different
models predict significantly different results. In the present work, we employed a coalescence
model (kernel) suggested by
\citeauthor{Coula1977} (\citeyear{Coula1977}):
\begin{linenomath}
\begin{equation}
    Q_{j,k} = \alpha(d_j,d_k) \omega(d_j,d_k)
\end{equation}
\end{linenomath}
where $\alpha(d_j,d_k)$ is the coalescence efficiency, and $\omega(d_j,d_k)$ is the collision frequency of randomly fluctuating spheres that is calculated by a well-known equation as (e.g. \citeauthor{Coula1977}, \citeyear{Coula1977}):
\begin{linenomath}
\begin{align}
    \omega(d_j,d_k) = C_1(d_j+d_k)^2\epsilon^{1/3}(d_j^{2/3} + d_k^{2/3})^{1/2}
\end{align}
\end{linenomath}
where $C_1$ is the model parameter.\\
\citeauthor{Coula1977} (\citeyear{Coula1977}) suggested the following equation for the coalescence
efficiency: 
\begin{linenomath}
\begin{align}\label{eq:coaleff}
    \alpha(d_j,d_k) = \exp\left(-C_2 \dfrac{\mu_f \rho_f \epsilon}{\gamma^2}\left(\dfrac{d_jd_k}{d_j+d_k}\right)^4 \right)
\end{align}
\end{linenomath}
where $C_2$ is the model parameter.\\
The coalescence model parameters, $C_1$ and $C_2$, identified under specific conditions can be found in the literature (e.g. \citeauthor{Laakkonen2006}, \citeyear{Laakkonen2006}). Note, the parameter $C_2$ in Eq.~\eqref{eq:coaleff} is dimensional.\\
Since unique values of the model coalescence parameters do not exist, they usually need to be tuned to fit experimental data. In our further analysis we assumed $C_1 = 1$ that is close to the value 0.88 suggested by \citeauthor{Laakkonen2006} (\citeyear{Laakkonen2006}), whereas the parameter $C_2$ was used as a tunable parameter to fit the experimental data. The parameter $C_2$ was chosen to be tunable because experiments, employed for model validation, were conducted for a brine-liquid $CO_2$ system, where coalescence rate was expected to be significantly lower than in the air-water system studied by \citeauthor{Laakkonen2006} (\citeyear{Laakkonen2006}).

\subsection{Dissolution term}
The droplet-fluid mass transfer (dissolution) process is a key phenomenon defining behavior of soluble droplets in a turbulent flow.\\
We calculated the derivative, expressing the dissolution term in the population balance equation as follows:
\begin{linenomath}
\begin{equation}
    \left(\dfrac{\partial N_i}{\partial t}\right)_{diss} = \dfrac{N_{i+1}}{x_{i+1} - x_i} \bigg| \dfrac{d x_{i+1}}{d t}\bigg|_{diss} - \dfrac{N_i}{x_i-x_{i-1}} \bigg| \dfrac{d x_i}{d t}\bigg|_{diss} \label{eq:diss}
\end{equation}
\end{linenomath}
where $(d{x_i}/d t)_{diss}$ is the rate of size change of a droplet of the $i$-th size fraction only due to dissolution.\\
This equation is obtained from mass balance formulated for a droplet of the $i$-th size fraction and its neighbouring size fractions during dissolution. The derivation of Eq.~\eqref{eq:diss} is given in the Appendix. The following limitation is imposed on application of Eq.~\eqref{eq:diss}: a volume reduction of a droplet of the size fraction $i$ during a single time step should be smaller than a difference of volumes of droplets of the fractions $i$ and $i-1$ respectively.

\noindent
The dissolution rate for the $i$-th size fraction droplet is calculated as follows
\begin{linenomath}
\begin{align}
    \left(\dfrac{dx_i}{d t}\right)_{diss} = -\dfrac{k\pi^{1/3}(6x_i)^{2/3}}{\rho_d}(C_s - C_{\infty})
\end{align}
\end{linenomath}
where $C_s$ is the saturation concentration of carbon dioxide in a bulk water fluid, $C_{\infty}$ is the concentration of dissolved carbon dioxide in water, $k$ is the mass transfer coefficient, $\rho_d$ is the droplet density.

\noindent
The mass transfer coefficient is determined as: 
\begin{linenomath}
\begin{align}
    k = \dfrac{ShD_{CO_2}}{d}
\end{align}
\end{linenomath}
where $D_{CO_2}$ is the molecular diffusivity of carbon dioxide in water and $Sh$ is the Sherwood
number.\\
The Sherwood number for a droplet transported in a turbulent pipe flow is calculated by an empirical correlation as follows (\citeauthor{Kress}, \citeyear{Kress}):
\begin{linenomath}
\begin{align}
    Sh = 0.34 \left(\dfrac{d_p}{D} \right)^2 Re^{0.94}Sc^{0.5}
\end{align}
\end{linenomath}
where $Sc = \nu_f/D$ is Schmidt number.\\
Note that both the concentration and saturation concentration of a dissolved gas in bulk water change
along a pipeline. At a moderate pressure, the saturation concentration is calculated by
Henry's law  (e.g. \citeauthor{Bird}, \citeyear{Bird}), according to which the saturation gas concentration is proportional to the pressure. The dissolved gas concentration in bulk fluid is calculated from the mass conservation for a gas phase. If at the initial time moment, no gas is dissolved in a liquid, the dissolved gas concentration evolution with a decrease in a droplet holdup is calculated as:
\begin{linenomath}
\begin{align}
    C_{\infty} = \dfrac{\rho_d (\phi_0 - \phi)}{1-\phi}
\end{align}
\end{linenomath}
where $\phi = \displaystyle \sum_{i=1}^M N_ix_i$ is the volume concentration of a dispersed phase, $\phi_0$ is the volume concentration of a dispersed phase at the initial time moment.

\section{Results and discussion}
\begin{figure}[h!]
\centering
\includegraphics[width = .8\textwidth]{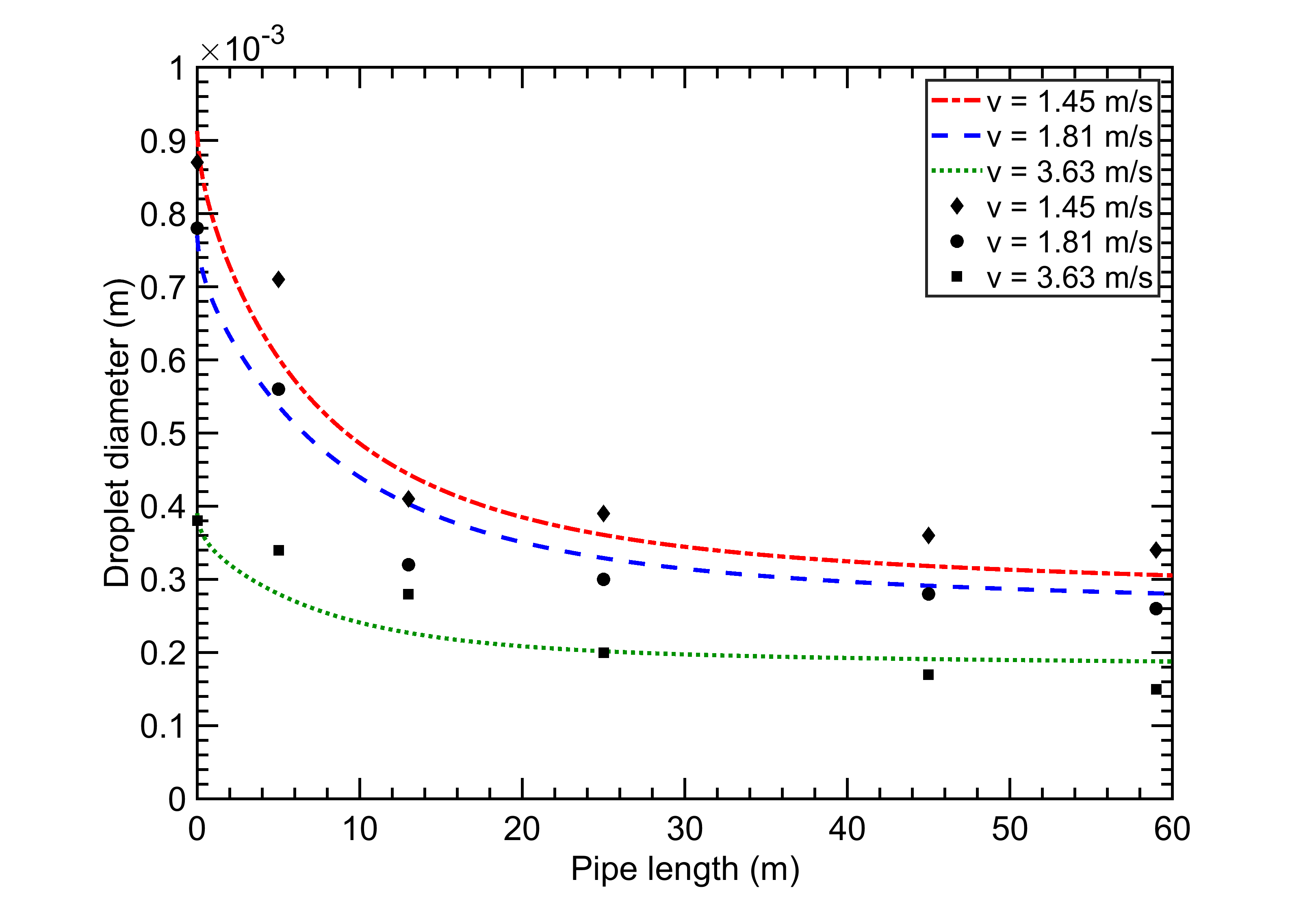}
  \caption{Distributions of computed droplet Sauter diameters along a pipe at different flow velocities versus the experimental data of  \citeauthor{Zendehboudi2013} (\citeyear{Zendehboudi2013}).}
  \label{fig:Figflowvel}
\end{figure}

The advection-diffusion population balance equation Eq.~\eqref{eq:1DTD} has been solved numerically. 
A MATLAB code was developed for this purpose. It is to be noted that in the simulations we have seen that calculations with the time steps of $10^{-4}$ and $10^{-5}$ provided virtually indistinguishable results, and thus the calculations are independent of the grid size. We also found that a further reduction in the radial mesh size and an increase in a number of droplet size fractions do not noticeably affect the computational results either. The code was validated against the experimental data given in \citeauthor{Zendehboudi2013} (\citeyear{Zendehboudi2013}). \citeauthor{Zendehboudi2013} (\citeyear{Zendehboudi2013}) measured changes in Sauter diameter along the pipe length for the volume fraction of $0.05$ under the following conditions: Pressure is $70$ bar, temperature is $25 ^{\circ} C$ and brine flow rate in the range $0.008 - 0.064$ m$^3$/s. At these conditions the solubility of $CO_2$ is approximately $5\%$. The solubility
decreases with temperature and increases with pressure. The reservoir temperature is higher than that used in the experiment, but the pressure is also higher; therefore, the solubility value used for modeling is in agreement with possible reservoir conditions. When designing real injection systems, the maximum dissolved $CO_2$ amount should never exceed the solubility at reservoir conditions to prevent escape of carbon dioxide from brine. In the experiments, the liquid $CO_2$ was merged with a flow of brine phase. The droplets were recorded and tracked using high-speed cameras. 
In the present work, for calculation of the $CO_2$ concentration we employed the
same assumption that \citeauthor{Zendehboudi2013} (\citeyear{Zendehboudi2013}) used in their calculations: The saturation
concentration in brine was evaluated as $Cs = 0.85$ $Cs$ (\text{in pure water}). This assumption provides reasonable data for $Cs$ at temperatures in the range of $20 - 100 ^{\circ} C$ and
pressures in the range of $0 - 80$ bar for salinities of $0.5 - 1.4$ mol/kg.
The computed Sauter diameter distributions along a pipe for different mean flow
velocities at the initial mean droplet concentration $\phi_0 = 0.05$ were matched to the measured
data by tuning the parameters $K$ in Eq.~\eqref{eq:G_d} describing droplet breakup rate and $C_2$ in Eq.~\eqref{eq:coaleff} defining the coalescence probability. As it was discussed above, the parameter $K$ was
not reliably determined in the past, whereas the parameter $C_2$ depends on the chemical composition of fluids composing a dispersion. The best fitting was obtained at $K=0.1$ and $C_2 = 10^{13}$.
One can see that the computational results correlate well with the measured data in Fig.~\ref{fig:Figflowvel}. The fitting is not very accurate along the first half of the pipe length, whereas the second half is characterized by a closer fit. These observations are primarily explained by experimental inaccuracy. The droplet size distributions along the initial pipe section are wide and rapidly
changed; therefore, in the experiment, the analysis of images obtained by using high speed cameras does not allow a highly accurate evaluation of the droplet Sauter diameters. The second half of the pipe is characterized with nearly steady-state droplet size distributions, which are relatively narrow, and therefore, the Sauter diameters were determined with a better fit than those in the first half of the pipe.

\begin{figure}[h!]
    \centering
    \includegraphics[width =  0.8\textwidth]{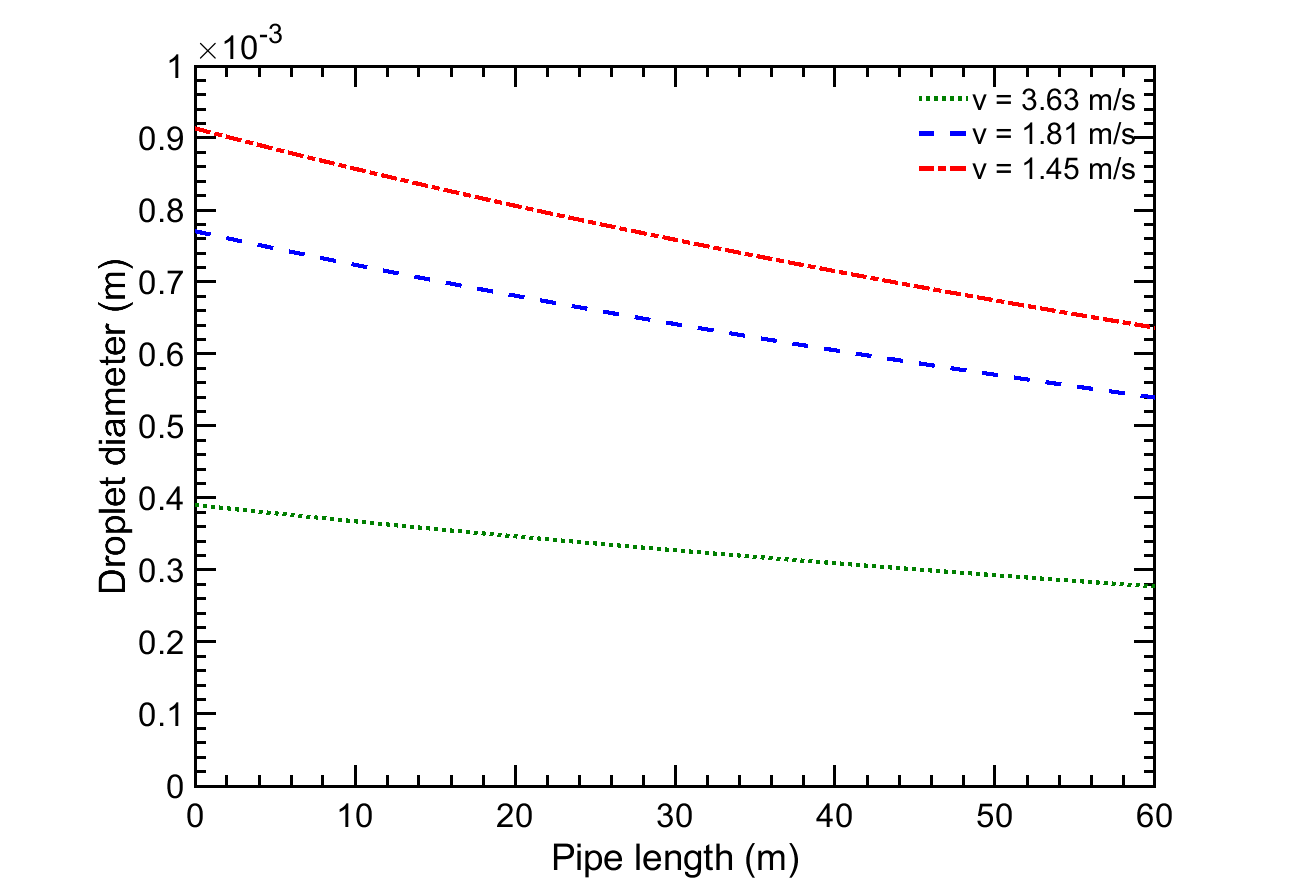}
    \caption{Evolution of droplet sizes along a pipe at different flow velocities accounting only for dissolution.}
    \label{fig:Fig_Diss}
\end{figure}

\noindent
Both droplet breakup and coalescence strongly affect the dissolution process. To illustrate the importance of these phenomena, in Fig.~\ref{fig:Fig_Diss} we showed the rate at which droplet sizes decrease along a pipe if breakup and coalescence are absent and droplet size is reduced only due to dissolution. 

\noindent
One can see that droplet sizes change slowly and do not approach steady-state.
The smaller the droplets, the higher the overall dissolution rate - which is, to a large extent, determined by specific droplet surface area.

\begin{figure}[h!]
    \centering
    \includegraphics[width=0.8\textwidth]{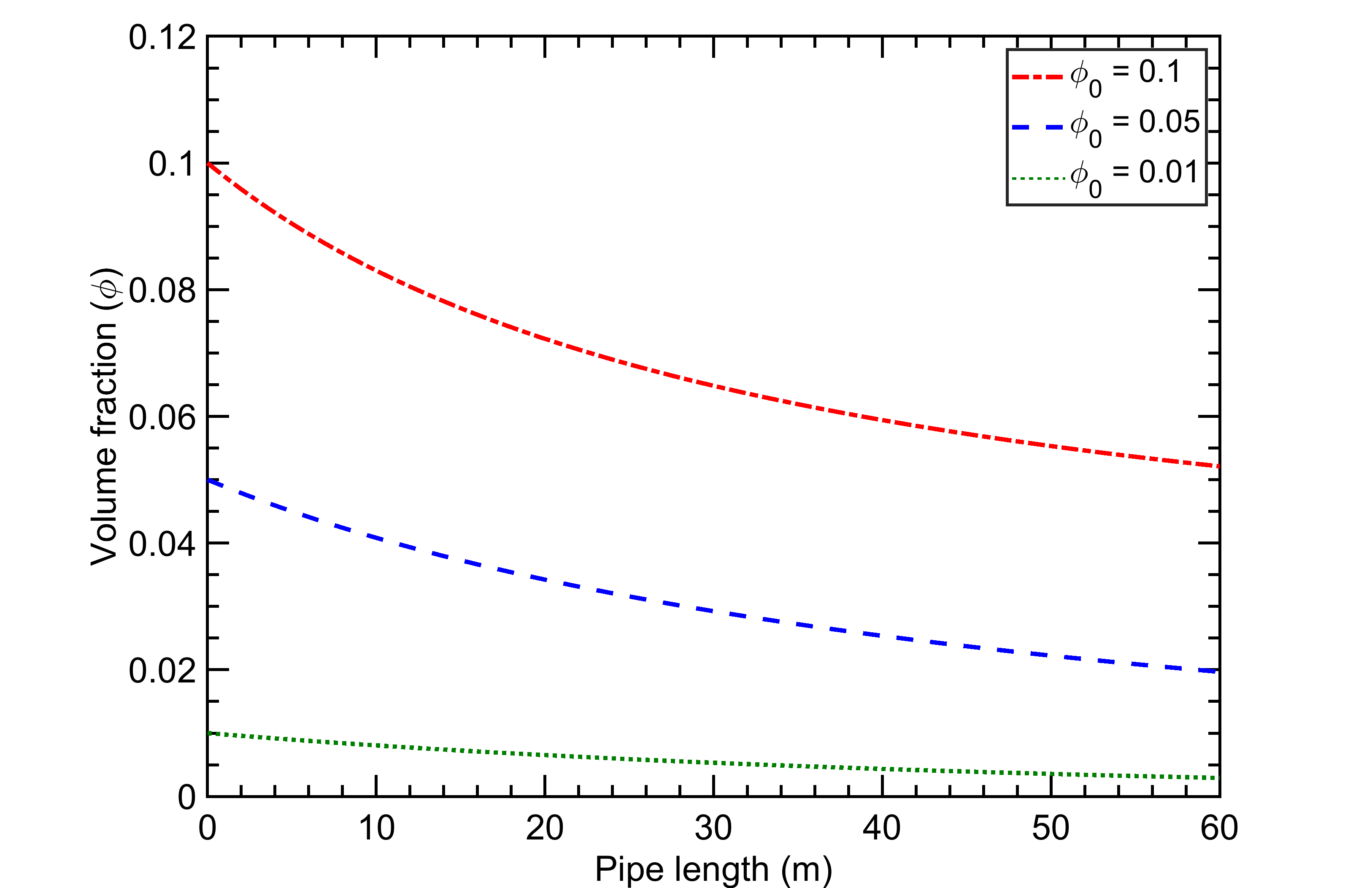}
    \caption{Evolution of droplet volume fractions along a pipe with different initial droplet volume concentrations at the fixed mean flow velocity $v = 3.63$ m/s.}
    \label{fig:Fig_evol_dvc}
\end{figure}

\begin{figure}[h!]
    \centering
    \includegraphics[width=0.8\textwidth]{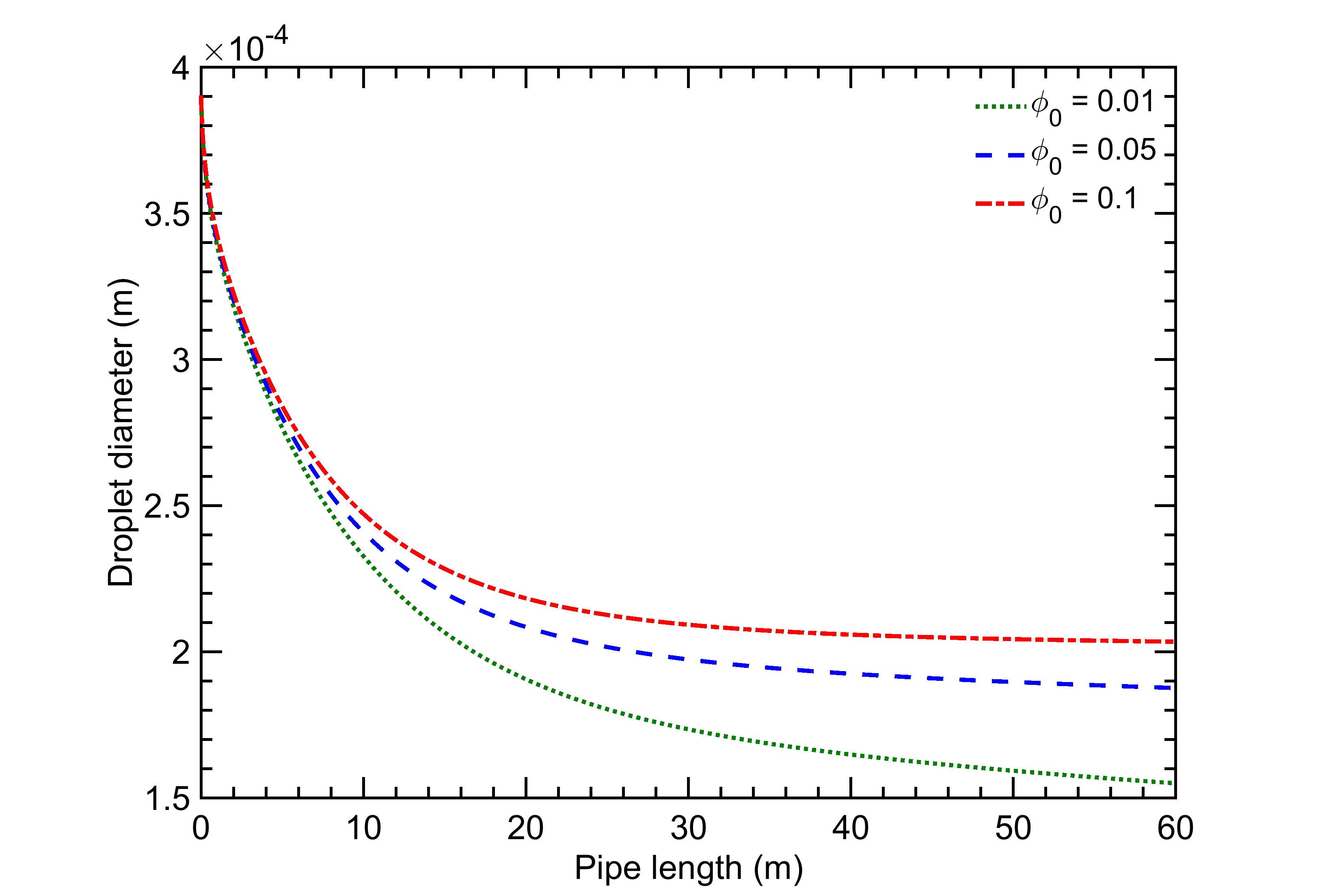}
    \caption{Evolution of droplet sizes along a pipe for different initial droplet volume concentrations at a fixed mean flow velocity of $v = 3.63$ m/s.}
    \label{fig:Fig_evol_sizes}
\end{figure}

\noindent
In Figs. \ref{fig:Fig_evol_dvc} and \ref{fig:Fig_evol_sizes} we illustrate an effect of the droplet volume fraction on the dissolution process. The calculations were conducted at the three different initial volume fractions $\phi_0 = 0.01,0.05,0.1$. The mean flow velocity was assumed to be the same for all the computations, $v=3.63$ m/s. Fig.~\ref{fig:Fig_evol_dvc} shows how the droplet volume fraction changes along the pipe. In Fig.~\ref{fig:Fig_evol_sizes}, one can see the Sauter diameter evolution.
The higher the droplet concentration, the closer the Sauter diameter approaches steady-state. This observation is explained as follows: A higher droplet concentration leads to a larger mass flux (from the dispersed to the continuous phase) which causes a faster increase in dissolved $CO_2$ concentration. This leads to the dissolved gas concentration rapidly approaching the saturation concentration, resulting in a slower mass transfer rate.

\begin{figure}[h!]
    \centering
    \includegraphics[width=0.8\textwidth]{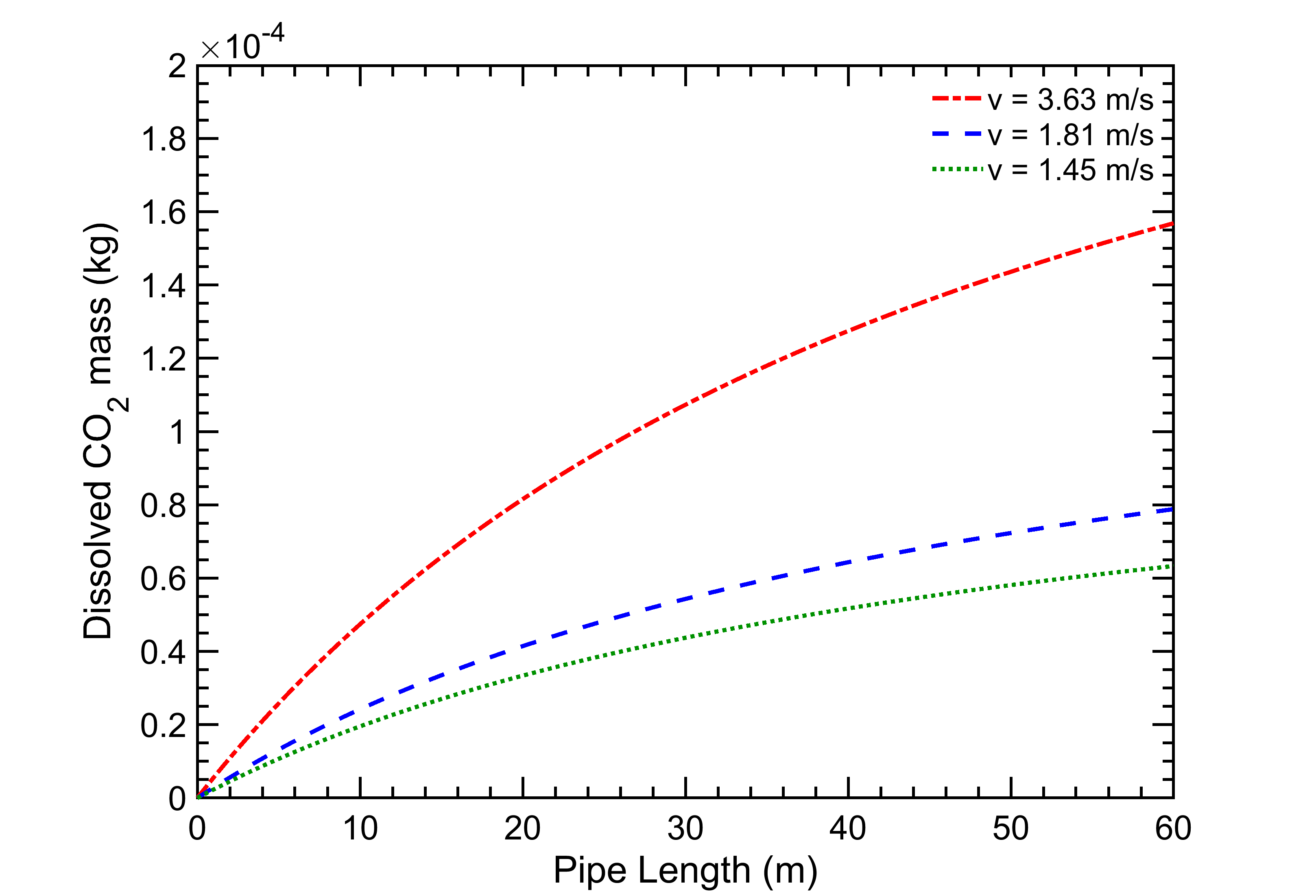}
    \caption{Evolution of dissolved $CO_2$ concentration by mass ($kg$) with different flow velocities at the fixed initial water content $\phi_0 = 0.05$.}
    \label{fig:Fig_evol_phi}
\end{figure}

\newpage
\noindent
In Fig.~\ref{fig:Fig_evol_phi} one can see an effect of the flow velocity on the change in the dissolved gas concentration along a pipe at the fixed initial droplet concentration $\phi_0 = 0.05$. The higher the flow velocity, the higher the mass transfer rate, resulting in a higher dissolved $CO_2$ concentration in water. This observation is explained by the following factors: 1) the higher the Reynolds number, the higher the mass transfer coefficient between a droplet and a surrounding liquid; and 2) the higher the flow velocity, the smaller the droplets - this leads to an increase in the droplet surface area.

\noindent
Based on the numerical study conducted, it is possible to conclude that the process of $CO_2$ dissolution in a pipe flow can be optimized by a proper selection of parameters such as the droplet concentration and the flow velocity. Note, the pipe diameter is also an important parameter allowing to vary the flow velocity if the flow rate is constrained. Overall, the dissolution process optimization is a relatively simple problem and the code presented here makes the optimization a rather straightforwardly handled task.  
Although the model developed is based on first principles and has a significantly higher predictive capability, it is capable of running relatively quickly on a regular laptop. Therefore, it is possible to formulate a simple approach, which will allow using the code developed to design an optimum dissolution process with a minimum time expense. Let us describe this approach.

\begin{figure}[h!]
    \centering
    \includegraphics[width=0.8\textwidth]{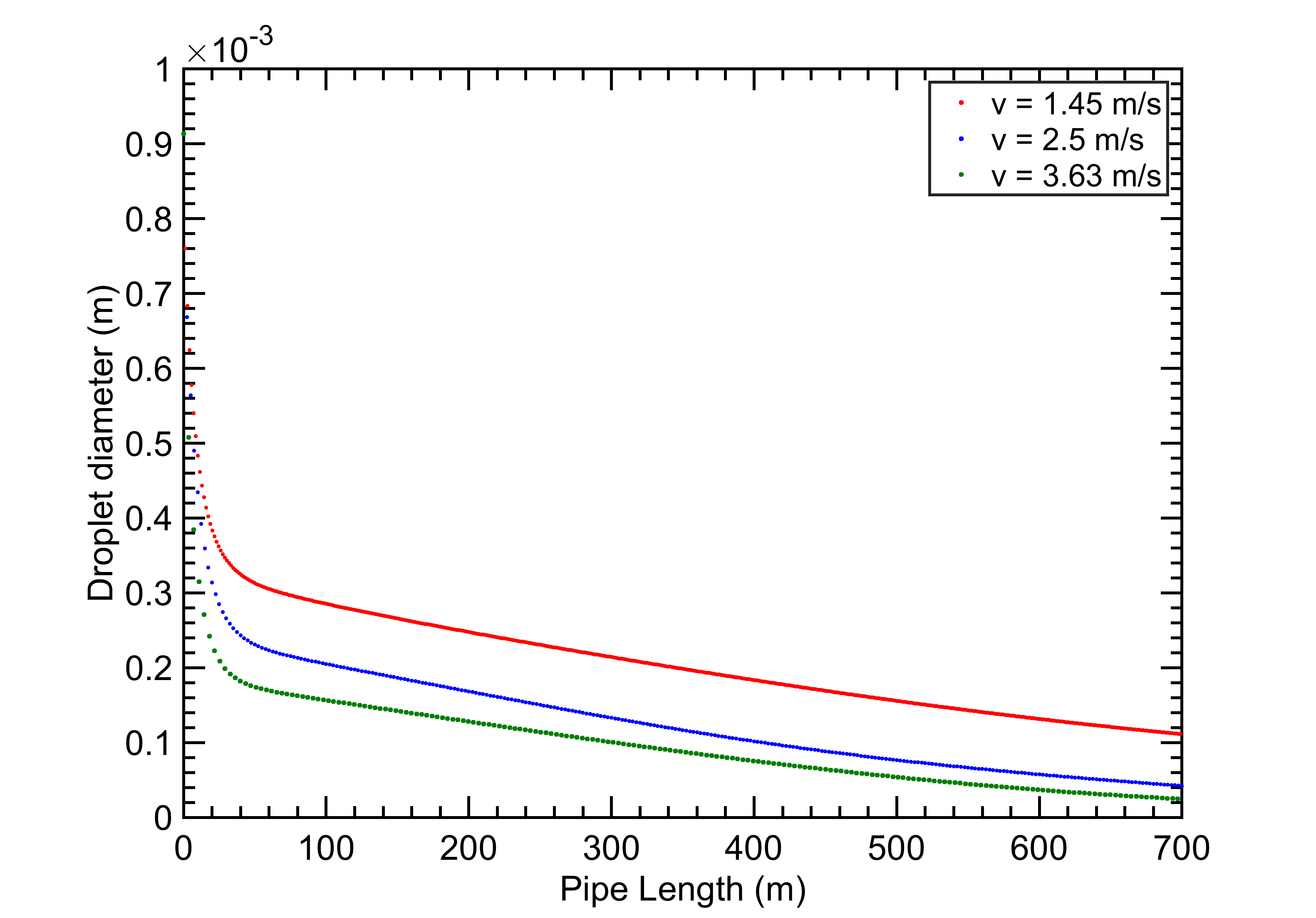}
    \caption{Evolution of droplet sizes along a pipe for different flow velocities with corresponding initial water contents.}
    \label{fig:evol_size_700}
\end{figure}

In engineering applications, a $CO_2$ flow rate is known. Pipe diameter is not, usually, varied significantly. Flow rate, determining $CO_2$ droplet concentration, is also restricted. Pipe length can usually be varied to a certain extent. Thus, the optimal combination of a water flow rate and a pipe length can be determined by a few computations. The following calculation strategy could be employed:
\begin{itemize}
    \item Conduct a computation using the most reasonable pipe length and the maximum possible flow rate. Results will show a degree of droplet dissolution.
    \item If both the droplet volume fraction and droplet sizes are sufficiently small, then further computations could be done at reduced flow rates. The smallest possible flow rate which allows to get satisfactory mixture parameters at the pipe end should be selected. 
    \item If a dissolution degree is not satisfactory, the only option available, in this case, is an increase in the pipe length. Otherwise, if an acceptable increase in the pipe length leads to acceptable droplet sizes, then the problem is solved. If the acceptable pipe length is not sufficient for dissolution, then a desired design is impossible to achieve. Maybe, using a system composed of two pipes could be employed.
\end{itemize}
To demonstrate the design process, we conducted computation in a pipe of $700$ m at a constant $CO_2$ flow rate corresponding to $\phi_0 = 0.05$ of $CO_2$ droplets at a mean flow velocity equal to $v = 1.45$ m/s. The other computations were done at the same $CO_2$ volume flow rate but for the higher mixture flow velocities, $v = 2.5$ m/s ($\phi_0 = 0.029$) and $v = 3.63$ m/s ($\phi_0 = 0.01997$), respectively. The computational results are shown in Fig.~\ref{fig:evol_size_700}. We can see that very fine droplet sizes are achievable and conclude that an increase in the brine flow rate along with an increase in the pipe length can cause substantial reduction in droplet sizes due to three main factors: 1) an increase in the flow velocity leads an increase in a turbulence energy dissipation rate level that in turn causes an increase in the droplet breakup rate; 2) the higher the flow velocity, the higher the droplet-fluid mass transfer coefficient; 3) the lower the droplet concentration, the higher the difference between the $CO_2$ saturation concentration and the concentration of $CO_2$ dissolved in a fluid causing a higher droplet-fluid mass transfer rate. 
Thus, in the majority of practical applications, a desired droplet size can be provided.

\newpage
\section{Conclusion}

\noindent
Thus, as a critical part of optimization of carbon dioxide underground disposal technology, a comprehensive model of ex-situ $CO_2$ droplet dissolution in a turbulent pipe water flow, accompanied with droplet breakup and coalescence, has been developed. Modeling has been reduced to a numerical solution of the steady-state advection-diffusion-population balance equation. The Prandtl mixing length model was employed for the modeling of the velocity distribution across a pipe. The turbulent diffusivity distribution along a pipe radius was calculated by the empirical correlation found in open literature. The turbulence energy dissipation rate, needed for population balance modelling, was calculated by the analytical equation. The population balance model developed accounts for droplet breakup, coalescence and turbulent mass transfer. Although both the breakup and coalescence terms of the discretized population balance equation were computed using known expressions, the dissolution term was derived in the present work. The semi-empirical models of droplet breakup, coalescence and dissolution were employed.\\
The model developed has been validated against the experimental data. The computed distributions of mean droplet Sauter diameter of liquid $CO_2$ along a pipe were compared with the measured distributions. The two model parameters were tuned for fitting the experimental results. The numerical studies, conducted by the validated model, showed that the droplet dissolution significantly speeds up with an increase in the flow velocity. An increase in the initial droplet concentration slows the dissolution process.\\
The model developed can be used for an efficient dissolution process design and optimization.

\section{Appendix}

\subsection{Dissolution}
The dissolution term of the population balance equation was derived by considering the mass balance for a droplet of volume $x_i$. The mass transfer rate is assumed to be described by
the function $\dfrac{dx_i}{dt}$.
We can formulate the mass balance equation between two particles of
volumes $x_i$ and $x_{i+1}$ due to dissolution during the time $dt$ as follows:
\begin{linenomath}
\begin{align}
\dfrac{\partial\hat{N}_{i+1}}{\partial t}x_i dt + N_{i+1}\dfrac{dx_{i+1}}{dt} dt = \dfrac{\partial\hat{N}_{i+1}}{\partial t}x_{i+1} dt \label{eq:dissappend}
\end{align}
\end{linenomath}
where $\dfrac{\partial \hat{N}_{i+1}}{\partial t}$
is the droplet number concentration rate of change of the size fraction $i+1$ only due to dissolution of droplets of this size fraction.
The first term is equal to the volume transferred from droplets of the volume $x_{i+1}$ to droplets of the
volume $x_i$ due to dissolution of droplets of the $i+1$ size fraction. The second term is equal to
the dissolved volume of droplets of the size fraction $i+1$. The right-hand side term is equal to the volume of droplets of the size fraction $i+1$ which disappeared due to dissolution. Note, the dissolution rate is assumed to be positive in our analysis; therefore, absolute value parentheses are used in Eq.~\eqref{eq:absvalue} and further. From Eq.~\eqref{eq:dissappend} we obtain:
\begin{linenomath}
\begin{align}
\dfrac{d\hat{N}_{i+1}}{dt} &= \dfrac{N_{i+1}}{x_{i+1}-x_i}\bigg|\dfrac{dx_{i+1}}{dt}\bigg| \label{eq:absvalue}
\end{align}
\end{linenomath}
Then, the volume flux from droplets of the size fraction $i+1$ to droplets of the size fraction $i$
due to dissolution is:
\begin{linenomath}
\begin{align}
q_i^+ = \dfrac{N_{i+1}x_i}{x_{i+1}-x_i}\bigg| \dfrac{dx_{i+1}}{dt} \bigg|, \label{eq:qi+}
\end{align}
\end{linenomath}
and the volume flux from droplets of the size fraction $i$ (indicating droplets disappearance due to dissolution) to droplets of the size fraction $i-1$ is as follows:
\begin{linenomath}
\begin{align}
q_i^- = \dfrac{N_ix_i}{x_i-x_{i-1}}\bigg| \dfrac{dx_i}{dt}\bigg|. \label{eq:qi-}
\end{align}
\end{linenomath}
Hence Eq.~\eqref{eq:qi+} and Eq.\eqref{eq:qi-} gives the total rate of number concentration change of $x_i$ due to dissolution,
\begin{linenomath}
\begin{equation}
\left(\dfrac{\partial N_i}{\partial t}\right)_{diss} = \dfrac{N_{i+1}}{x_{i+1}-x_i} \bigg| {\dfrac{dx_{i+1}}{dt}} \bigg|- \dfrac{N_i}{x_i - x_{i-1}} \bigg| {\dfrac{dx_i}{dt}} \bigg|.
\end{equation}
\end{linenomath}
For both the smallest and the largest size fractions this equation is modified as follows:
\begin{linenomath}
\begin{align}
\dfrac{\partial N_1}{\partial t} &= \dfrac{N_2}{x_2-x_1}\bigg| \dfrac{dx_2}{dt}\bigg| - \dfrac{N_1}{x_1}\bigg| \dfrac{dx_1}{dt}\bigg| \hspace{.5cm} \text{for the first size fraction $x_1$}\\
\dfrac{\partial N_m}{\partial t} &= -\dfrac{N_m}{x_m-x_{m-1}}\bigg| \dfrac{dx_m}{dt}\bigg| \hspace{.5cm} \text{for the last size fraction $x_m$}
\end{align}
\end{linenomath}

\newpage
\begin{acknowledgement}

Financial support for this work provided by Natural Sciences and Engineering
Research Council of Canada (NSERC).
\end{acknowledgement}\\

\noindent
Declaration of interest: None


\newpage
\bibliography{achemso-demo}

\begin{tocentry}
  \includegraphics[width=\linewidth]{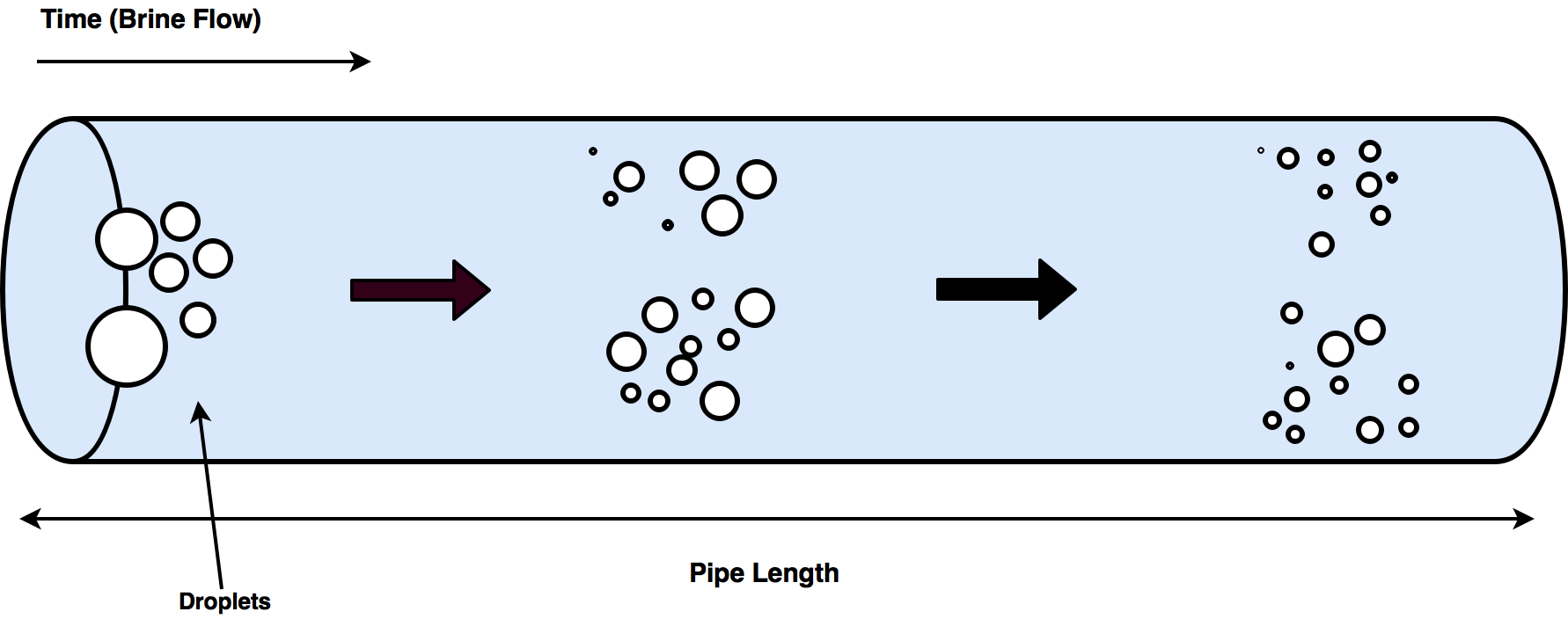}
\end{tocentry}

\end{document}